\documentclass[a4paper,11pt]{article}
\pdfoutput=1 

\usepackage{jcappub} 

\usepackage[T1]{fontenc} 

\usepackage{amsmath, amsfonts, amsthm, amssymb, graphicx,comment}

\usepackage{hyperref}

\allowdisplaybreaks[3]

\def\be{\begin{equation}}
\def\ee{\end{equation}}
\def\ba{\begin{eqnarray}}
\def\ea{\end{eqnarray}}
\def\ba{\begin{eqnarray}}
\def\ea{\end{eqnarray}}
\newcommand{\nn}{\nonumber\\}
\newcommand{\ud}{\mathrm{d}}
\def \pd {\partial}

\def\({\left(}
\def\){\right)}
\def\[{\left[}
\def\]{\right]}

\title{\huge Generalized massive gravity in arbitrary dimensions and its Hamiltonian formulation}

\author[a]{Qing-Guo Huang,}
\author[a,b]{Ke-Chao Zhang,}
\author[c]{Shuang-Yong Zhou}

\affiliation[a]{State Key Laboratory of Theoretical Physics, Institute of Theoretical Physics,
Chinese Academy of Sciences, P.O. Box 2735, Beijing 100190, China}
\affiliation[b]{University of Chinese Academy of Sciences, Beijing 100190, China}
\affiliation[c]{SISSA, Via Bonomea 265, 34136, Trieste, Italy and INFN Sezione di Trieste, Italy}

\emailAdd{huangqg@itp.ac.cn}
\emailAdd{zkc@itp.ac.cn}
\emailAdd{szhou@sissa.it}

\abstract{
We extend the four-dimensional de Rham-Gabadadze-Tolley (dRGT) massive gravity model to a general scalar massive-tensor theory in arbitrary dimensions, coupling a dRGT massive graviton to multiple scalars and allowing for generic kinetic and mass matrix mixing between the massive graviton and the scalars, and derive its Hamiltonian formulation and associated constraint system. When passing to the Hamiltonian formulation, two different sectors arise: a general sector and a special sector. Although obtained via different ways, there are two second class constraints in either of the two sectors, eliminating the BD ghost. However, for the special sector, there are still ghost instabilities except for the case of two dimensions. In particular, for the special sector with one scalar, there is a ``second BD ghost''.
}

\begin{document}

\maketitle
\flushbottom

\section{Introduction}

The linear Fierz-Pauli model \cite{Fierz:1939ix} is a unique linear massive-tensor theory in flat spacetime that is free of ghost instabilities, but phenomenologically this model differs unacceptably from linearized General Relativity, say, in the solar system, which is known as the van Dam-Veltman-Zakharov (vDVZ) discontinuity \cite{vanDam:1970vg, Zakharov:1970cc}. Nonlinear extensions of the linear Fierz-Pauli model generally cures the vDVZ discontinuity via the Vainshtein mechanism \cite{Vainshtein:1972sx, Babichev:2013usa}, but they also typically contain the nonlinear Boulware-Deser (BD) ghost \cite{Boulware:1973my}. Recently, a family of nonlinear massive gravity models, de Rham-Gabadadze-Tolley (dRGT) massive gravity, has been discovered \cite{deRham:2010kj, deRham:2010ik,deRham:2010gu} and shown to be free of the BD ghost \cite{Hassan:2011hr,Hassan:2011ea, deRham:2010kj, Hassan:2011tf} (see also \cite{deRham:2011rn,Hassan:2012qv,deRham:2011qq,Kluson:2012wf,Mirbabayi:2011aa,Golovnev:2011aa,Alexandrov:2012yv}). See \cite{Hinterbichler:2011tt} for a recent review of massive gravity.

One motivation of the recent re-consideration of massive gravity is that it may account for the late time cosmic acceleration if the graviton mass is of the Hubble scale. Indeed, self-accelerating solutions have been found in dRGT massive gravity \cite{D'Amico:2011jj,Gumrukcuoglu:2011ew, Volkov:2012cf, Volkov:2012zb, Gratia:2012wt,Koyama:2011xz,deRham:2010tw,Chamseddine:2011bu,Kobayashi:2012fz,Nieuwenhuizen:2011sq, Berezhiani:2011mt}.  While the dRGT model with a flat fiducial metric does not admit a spatially flat or closed Friedmann-Robertson-Walker (FRW) solution that is fully homogeneous and isotropic \cite{D'Amico:2011jj}, the open FRW solution does exist \cite{Gumrukcuoglu:2011ew}. However, around this FRW solution, the scalar and vector modes of the linear perturbations are found to have vanishing kinetic terms \cite{Gumrukcuoglu:2011zh}, and nonlinearly this FRW solution has been shown to suffer from a ghost instability \cite{DeFelice:2012mx}.  The instability problem for the self-accelerating solutions seems to persist even if the homogeneity and isotropy requirement is relaxed for the non-metric components in the Stuckelberg language \cite{Koyama:2011wx, Tasinato:2012ze} or if a more general fiducial metric is considered  \cite{DeFelice:2012mx, DeFelice:2013awa} (see also \cite{Fasiello:2012rw} for a non-self-accelerating case).  However, see \cite{Gumrukcuoglu:2012aa, DeFelice:2013awa} for potentially healthy anisotropic solutions. In any case, the anisotropic configurations in the Stuckelberg fields reflect physical anisotropies, as may be more evident in unitary gauge.

In massive gravity, diffeomorphism invariance is explicitly broken by the graviton potential. In field theories, a gauge invariance is often broken spontaneously. A simple way to formulate a spontaneous breaking in the dRGT model is to promote its free parameters to depend on an extra scalar field \cite{D'Amico:2011jj, Huang:2012pe} (mass-varying massive gravity), and it has been shown that this promotion does not re-introduce the BD ghost \cite{Huang:2012pe}. This extension turns out to have rich cosmological implications \cite{Huang:2012pe, D'Amico:2011jj, Saridakis:2012jy, Cai:2012ag, Wu:2013ii, Hinterbichler:2013dv, Gumrukcuoglu:2013nza}. After coupling to an extra scalar, a flat FRW universe becomes a solution \cite{D'Amico:2011jj, Huang:2012pe}, and, more importantly, all the modes of the FRW solution's linear perturbations can now have non-vanishing kinetic terms, and parameter space without ghost instabilities exists, provided a few conditions are satisfied \cite{Gumrukcuoglu:2013nza}.  A couple of other ways to couple scalars to a dRGT graviton have been proposed that are also free of the BD ghost \cite{D'Amico:2012zv,Gannouji:2013rwa, Hinterbichler:2013dv, Andrews:2013ora}. However, not all BD ghost-free scalar extensions of the dRGT model can have a stable FRW solution \cite{Gumrukcuoglu:2013nza}.

In this paper, we further generalize mass-varying massive gravity to a general scalar massive-tensor theory and derive its Hamiltonian formulation. In this general setup, the massive graviton couples to multiple scalars, which may or may not form a symmetry group representation, and generic kinetic mixing and mass matrix mixing between the massive graviton and the scalars are considered. These mixings, which have not been considered before, may extend possible parameter space of stable FRW solutions for a dGRT-like theory, enlarging the theory space for further consistency and phenomenological checks. Particularly, it would be interesting to see how the results of \cite{Gumrukcuoglu:2013nza} will be enhanced by taking into account the these kinetic and mass mixings. 

We also formulate the model in arbitrary dimensions, as there may be applications of dRGT-like gravity in higher or lower than four dimensions. For example, a Kaluza-Klein reduction of a higher dimensional gravity theory usually leads to massive gravitons coupled to additional scalar modes, and the lower dimensional theory may not be necessarily four dimensional. We will see that after passing to the Hamiltonian formulation two different sectors arise, whose Hamiltonians and constraints are to be derived separately. In the general sector (see the definition immediately before Section \ref{sec:consys}) there is one primary and one secondary constraint, while in the special sector there are two primary constraints but no secondary constraint. Thus, the BD ghost is eliminated in both of them. However, for the special sector, there are ghost instabilities except for the case of two dimensional spacetime. In particular, the special sector with one scalar is unstable due to a ``second BD ghost''.  An explicit Hamiltonian formulation will presumably be useful to compute energies of gravitational solutions in various dimensions, to numerically evolve field configurations in time and to even discuss possible canonical quantization of the theory.

This paper is organized as follows. In Section \ref{sec:action}, we introduce our theory in $d+1$ dimensions. In Section \ref{sec:ADM}, we apply a $d+1$ decomposition to the action and show that two different sectors arise when passing to the Hamiltonian formulation. In Section \ref{sec:consys}, we derive the Hamiltonians and their constraints for the two sectors separately. The discussion of the special sector will be less detailed, as there are overlaps with the general sector. Nevertheless, this calculation is necessary to show that there is a ``second BD ghost'' for the special sector with one scalar and that the two dimensional case is free of the BD ghost. Some detailed calculations in Section \ref{sec:ADM} and \ref{sec:consys} are put in the Appendices to make the presentation concise. We conclude and outlook in Section \ref{sec:conclu}.

\section{The action} \label{sec:action}

As mentioned in the Introduction, the fully homogeneous and isotropic solutions in the dRGT model suffer from a ghost instability \cite{DeFelice:2012mx}. By adding extra degrees of freedom, this problem may be avoided \cite{Gumrukcuoglu:2013nza} in some extensions of the dRGT model that are also free of the BD ghost, such as mass-varying massive gravity \cite{Huang:2012pe}, provided a few conditions are satisfied \cite{Gumrukcuoglu:2013nza}. Here we consider a further generalization of mass-varying massive gravity, which may provide further theory space to obtain a phenomenologically viable homogeneous and isotropic solution.

Consider a general scalar massive-tensor theory where $\mathcal{N}$ scalars $\phi_A$ ($A=1,2,...,\mathcal{N}$) are non-minimally coupled to a massive graviton whose potential interactions are organized in the dGRT form \cite{deRham:2010kj}. Our specification for the field space of the $\mathcal{N}$ scalars are left general, so $\phi_A$ may or may not have (global) internal symmetry. We also formulate the theory in arbitrary dimensions, as potentially there will be applications of massive gravity in higher or lower than four dimensions. The unitary gauge action in $d+1$ dimensions is given by
\be \label{startaction0}
S = \int \ud^{d+1} x ~\sqrt{-g}\left[ \Omega(\phi_A) \frac{R}{2} + \sum_{a=1}^{d+1}\alpha_a(\phi_A)e_a(\mathcal{K})
-\frac12 \pd_\mu \phi_A \pd^\mu \phi^A - V(\phi_A) \right]     ,
\ee
where $e_a(\mathcal{K})$ takes the dRGT form
\begin{align}
e_a(\mathcal{K})  =  \mathcal{K}^{\mu_1}_{[\mu_1}\mathcal{K}^{\mu_2}_{\mu_2} \cdots\mathcal{K}^{\mu_a}_{\mu_a]}   ,
\qquad
\mathcal{K}^\mu_\nu = \delta^\mu_\nu-\sqrt{g^{\mu\rho}f_{\rho\nu} }     ,
\end{align}
all the spacetime indices are raised or lowered with $g_{\mu\nu}$, and $f_{\mu\nu}$ is a general fiducial metric which does not have dynamics. $\Omega(\phi_A)$, $\alpha_a(\phi_A)$ and $V(\phi_A)$ are general functions of $\phi_A$ or invariants of $\phi_A$ in the case of an internal symmetry.  $\sqrt{g^{\mu\rho}f_{\rho\nu} }$ is defined such that $g^{\mu\rho}f_{\rho\nu}  = \sqrt{g^{\mu\rho}f_{\rho\sigma}} \sqrt{g^{\sigma\lambda}f_{\lambda\nu}}$ (or in matrix form ${\rm g}^{-1}f=\sqrt{{\rm g}^{-1}f}\sqrt{{\rm g}^{-1}f}$), and the antisymmetrization is defined by $\mathcal{K}^\mu_{[\mu} \mathcal{K}^\nu_{\nu]} =(\mathcal{K}^\mu_{\mu} \mathcal{K}^\nu_{\nu} -\mathcal{K}^\mu_{\nu}\mathcal{K}^\nu_{\mu})/2!$ and so on. Also, we have chosen the reduced Planck mass $M_P$ to be 1, and summation over repeated scalar labels is assumed.

Note that by assigning $\Omega(\phi_A)$ we allow the scalars and the massive graviton to be kinetically mixed. One may diagonalize their kinetic terms by a conformal transformation, but matter may minimally couple to the gravity sector in the Jordan frame, so once matter fields are added they give rise to different dynamics. We have also allowed $a=1$ in action (\ref{startaction0}), which has been neglected in previous attempts to extend the dRGT model. (Note that in the pure dRGT model the corresponding $a=1$ term gives rise to a tadpole contribution for the graviton around the fiducial metric and therefore is not usually considered.)  Adding this term means that we allow non-zero values for the off-diagonal components of the scalars and the massive graviton's mass matrix. If we assume there is no tadpole term for the graviton, $\alpha_1$ should be chosen to be of the same order of the perturbative metric around the fiducial metric. In the Stuckelberg picture, derivatives of the Stuckelberg modes arise in the graviton potential, so the $a=1$ term also in a sense adds extra kinetic mixing. It would be interesting to see how these new ingredients generalize the theory space for healthy FRW solutions. However, in this paper, we will focus on eliminating obviously problematic cases of the general theory (\ref{startaction0}), including cases that suffer from the BD ghost.

To show whether the BD ghost is present or not, we will switch to Hamiltonian formulation, for which it is more convenient to cast the action in terms of~\cite{Hassan:2011vm}
\begin{align}
e_a(\mathcal{X})  =  \mathcal{X}^{\mu_1}_{[\mu_1}\mathcal{X}^{\mu_2}_{\mu_2} \cdots\mathcal{X}^{\mu_a}_{\mu_a]}    ,
\qquad
\mathcal{X}^\mu_\nu = \sqrt{g^{\mu\rho}f_{\rho\nu} }     ,
\end{align}
so that we have the action
\be \label{actionstart}
S = \int \ud^{d+1} x ~\sqrt{-g}\left[ \Omega(\phi_A) \frac{R}{2} + \sum_{a=0}^{d+1}\beta_a(\phi_A)e_a(\mathcal{X})
-\frac12 \pd_\mu \phi_A \pd^\mu \phi^A - V(\phi_A) \right]      ,
\ee
where $\alpha_n(\phi_A)$ and $\beta_n(\phi_A)$ are related by
\begin{align}
\beta_n = (-1)^n \sum_{a=n}^{d+1} \frac{(d+1-n)!}{(d+1-a)!(a-n)!}\alpha_a , \quad
\alpha_0 \equiv 0 .
\end{align}
Note that in the dRGT model the term $\beta_{d+1}e_{d+1}(\mathcal{X})=\beta_{d+1}\det({\rm g}^{-1}f)^{\frac12}$ ($\beta_{d+1}$ being constant) is not dynamical and thus can be dropped, but here $\beta_{d+1}=\beta_{d+1}(\phi_A)$, so it is generally dynamical.

\section{ADM decomposition} \label{sec:ADM}

To reformulate the theory (\ref{actionstart}) in the Hamiltonian form, we first map the spacetime with ADM coordinates and write the two metrics as
\begin{align}
g_{\mu\nu}\ud x^\mu \ud x^\nu &= - N^2\ud t^2 +\gamma_{ij}(\ud x^i+N^i\ud t)(\ud x^j+N^j\ud t) ,
 \\
f_{\mu\nu}\ud x^\mu \ud x^\nu &= - L^2\ud t^2 +\xi_{ij}(\ud x^i+L^i\ud t)(\ud x^j+L^j\ud t)  ,
\end{align}
where $\gamma_{ij}$ and $\xi_{ij}$ are induced metrics on the $d$-dimensional hypersurface and $N$, $N^i$, $L$ and $L^i$ are lapses and shifts with respect to $g_{\mu\nu}$ and $f_{\mu\nu}$ respectively. Since $\xi_{ij}$ is assumed to be fixed,  only the extrinsic curvature associated with $g_{\mu\nu}$ is relevant for the Hamiltonian formulation, which will be written as $K_{ij}=\left( \dot{\gamma}_{ij}- 2D_{(i}N_{j)} \right)/2N$, where $D_i$ is the covariant derivative associated with ${\gamma}_{ij}$ on the space-like hypersurface. In what follows, all $d$-dimensional indices are raised and lowered by $\gamma_{ij}$, and $x$ in a general function $f(x)$ only refers to the spatial coordinates.

Making use of the Gauss-Codacci relations and the expression $\bot^\mu=(1/N,-N^i/N)$ for the unit vector normal to the hypersurface (associated with $g_{\mu\nu}$) and neglecting some boundary terms, the curvature part and the scalar part of the Lagrangian can be decomposed respectively as
\begin{align}
\mathcal{L}_R
&=\sqrt{-g} \Omega \frac{R}{2}
\nn
&= N\sqrt{\gamma} \left[\frac12\Omega({}^{(d)\!}R + K_{ij}K^{ij} - K^2 )  -    D_i D^i\Omega \right]  - \sqrt{\gamma} K\Omega^{,A} ( \dot{\phi}_A -N^i \pd_i \phi_A  ) ,
\\
\mathcal{L}_S
&=\sqrt{-g}\left(-\frac12 \pd_\mu \phi_A \pd^\mu \phi^A - V\right)
\nn
& = \frac{\sqrt{\gamma}}{2N} \dot{\phi}_A \dot{\phi}^A - \frac{\sqrt{\gamma}}{N}\dot{\phi}_A N^i\pd_i \phi^A -\frac{N\sqrt{\gamma}}{2}  \pd_i \phi_A \pd^i \phi^A + \frac{\sqrt{\gamma}}{2N} N^i\pd_i \phi_AN^j\pd_j \phi^A - N\sqrt{\gamma} V ,
\end{align}
where ${}^{(d)\!}R$ is the Ricci scalar on the $d$-dimensional hypersurface and $\Omega^{,A}\equiv \pd \Omega/\pd \phi_A$.

We also want to decompose the graviton potential part
\be
\mathcal{L}_{GP}  = \sqrt{-g} \left[ \sum_{a=0}^{d+1}\beta_a e_a(\mathcal{X}) \right]
\ee
according to the prescription of \cite{Hassan:2011hr}, so that it will be easy to obtain the constraint system of the model and prove there is no BD ghost in the next section. To this end, we introduce a new ``shift vector'' $n^j$ satisfying \cite{Hassan:2011hr, deRham:2010kj}
\be
N^i - L^i = (L\delta^i_j+N\mathcal{D}^i{}_j)n^j  ,
\ee
where $\mathcal{D}^i{}_j$ is defined by the following equation
\be \label{Ddefrel}
x \mathcal{D}^i{}_k \mathcal{D}^k{}_j = (\gamma^{ik}-\mathcal{D}^i{}_m n^m \mathcal{D}^k{}_l n^l)\xi_{kj}, \quad x \equiv 1-n^i\xi_{ij}n^j  .
\ee
The last equation can be solved to obtain
$\mathcal{D}^i{}_j=(x\delta^i_l+n^in^k\xi_{kl})^{-1}\sqrt{(x\delta^l_n+n^ln^m\xi_{mn})\gamma^{np}\xi_{pj}}$, but it is often more convenient to use Eq.~(\ref{Ddefrel}) when dealing with $\mathcal{D}^i{}_j$. We further introduce
\begin{align}
 A^\mu_\nu&={1\over \sqrt x} \left(
    \begin{array}{cc}
      L+n^lL_l & n^l\xi_{lj} \\
      -(L+n^lL_l)(Ln^i+L^i) & -(Ln^i+L^i)n^l\xi_{lj} \\
    \end{array}
  \right) ,
  \\
B^\mu_\nu&=\sqrt x\left(
    \begin{array}{cc}
      0 & 0 \\
      \mathcal{D}^i{}_k\xi^{kl}L_l & \mathcal{D}^i{}_j \\
    \end{array}
  \right) .
\end{align}
Notice that $A^\mu_\nu,B^\mu_\nu$ and $\mathcal{D}^i{}_j$ have following useful properties
\begin{align}
\label{ABrel}
AB^n AB^m &= [AB^n] AB^m, \quad m,n\geq 0
\\
\label{matrixeq2}
[B^n]&=(\sqrt{x})^n[\mathcal{D}^n],
 \\
 \label{matrixeq3}
[AB^n]&=-L(\sqrt{x})^{n-1}n^j\xi_{ji}(\mathcal{D}^n)^i{}_k n^k,
\\
\label{matrixeq4}
 \xi_{ij}(\mathcal{D}^n)^j{}_k &=\xi_{kj}(\mathcal{D}^n)^j{}_i ,
\end{align}
where $[~]$ means the trace of the matrix contained and $A^n$, for example, means the $n$-th power of the matrix $A$.

Now, with these definitions, $\mathcal{X}^\mu_\nu$ can be written  as
\be
{\mathcal{X}}^\mu_\nu  = \sqrt{g^{\mu\rho}f_{\rho\nu}}=\frac{1}{N}A^\mu_\nu+B^\mu_\nu
\ee
and then we have
\begin{align}\label{ea1}
e_a(\mathcal{X})  &=  ({1\over N}A+B)^{\mu_1}_{[\mu_1}({1\over N}A+B)^{\mu_2}_{\mu_2} \cdots({1\over N}A+B)^{\mu_a}_{\mu_a]} \nn
                  &=  \sum_{q=0}^a\frac{a!}{q!(a-q)!}{1\over N^q}A^{\mu_1}_{[\mu_1}\cdots A^{\mu_q}_{\mu_q}B^{\mu_{q+1}}_{\mu_{q+1}}\cdots B^{\mu_a}_{\mu_a]} .
\end{align}

We can prove that all the terms with $q>1$ in Eq.~(\ref{ea1}) vanish. Consider a generic term (or more accurately, a bunch of terms organized by the antisymmetrization) with $q$ of $A$ and $a-q$ of $B$ and expand the antisymmetry out.  Making use of the relation (\ref{ABrel}),  a generic term in the expansion can be written as
\begin{align} \label{ABgenericterm}
&~~~~ [AB^{n_1}][AB^{n_2}]\cdots[AB^{n_q}] [B^{n_{q+1}}]\cdots[B^{n_{q+p}}]
\\
&=
 A^{\mu_1}_{\nu_1}(B^{n_1})^{\nu_1}_{\mu_1}A^{\mu_2}_{\nu_2}(B^{n_2})^{\nu_2}_{\mu_2} \cdots A^{\mu_q}_{\nu_q}(B^{n_q})^{\nu_q}_{\mu_q}[B^{n_{q+1}}]\cdots[B^{n_{q+p}}],
\end{align}
where $n_i\geq 0$, $n_1+n_2+\cdots+n_{q+p}=a-q$, and we have restricted to the case where $q$ is greater than 1. Since originally in Eq.~(\ref{ea1}) the lower indices are antisymmetrized, there must be another term in the expansion given by exchanging $\nu_1$ and $\nu_2$ and multiplied by a minus sign
\begin{align}
&~~~ -A^{\mu_1}_{\nu_2}(B^{n_1})^{\nu_1}_{\mu_1}A^{\mu_2}_{\nu_1}(B^{n_2})^{\nu_2}_{\mu_2} \cdots A^{\mu_q}_{\nu_q}(B^{n_q})^{\nu_q}_{\mu_q} [B^{n_{q+1}}]\cdots[B^{n_{q+p}}]
\\
&=
-[AB^{n_1}AB^{n_2}]\cdots[AB^{n_q}] [B^{n_{q+1}}]\cdots[B^{n_{q+p}}]
\\
&= -[AB^{n_1}][AB^{n_2}]\cdots[AB^{n_q}] [B^{n_{q+1}}]\cdots[B^{n_{q+p}}].
\end{align}
So this term exactly cancels the term given by (\ref{ABgenericterm}). Since the term given by (\ref{ABgenericterm}) is generic, this cancellation happens for any terms with $q>1$, which means in Eq.~(\ref{ea1}) terms with $q>1$ vanish.

So we are left with the terms with $q=0,1$. Making use of Eqs.~(\ref{ABrel}-\ref{matrixeq4}), the term with $q=1$ can be decomposed as
\begin{align}
{a\over N}A^{\mu_1}_{[\mu_1}B^{\mu_2}_{\mu_2}\cdots B^{\mu_a}_{\mu_a]}
 &= {L\over N}\bigg[ (\sqrt x)^a\mathcal{D}^{p_1}{}_{[p_1}\cdots \mathcal{D}^{p_{a-1}}{}_{p_{a-1}]} \nn
 &+ \sum_{i=1}^{a-1}(-1)^{(i+1)} (\sqrt x)^{a-2}n^l\xi_{lj}(\mathcal{D}^i)^j{}_kn^k \mathcal{D}^{p_{i+1}}{}_{[p_{i+1}}\cdots \mathcal{D}^{p_{a-1}}{}_{p_{a-1}]}  \bigg]
\end{align}
and the term with $q=0$ can be decomposed as
\begin{align}
B^{\mu_1}_{[\mu_1}\cdots B^{\mu_a}_{\mu_a]} = (\sqrt x)^a\mathcal{D}^{p_1}{}_{[p_1}\cdots \mathcal{D}^{p_a}{}_{p_a]}.
\end{align}
So  we have
\be
\mathcal{L}_{GP} = L\sqrt{\gamma} \mathcal{A} + N\sqrt{\gamma} \mathcal{B},
\ee
where $\mathcal{A}$ and $\mathcal{B}$ are given by
\begin{align}\label{CAdef}
\mathcal{A}  =\;& \sum_{a=1}^{d}\beta_a\bigg[ \sum_{i=1}^{a-1}(-1)^{(i+1)} (\sqrt x)^{a-2}n^l\xi_{lj}(\mathcal{D}^i)^j{}_kn^k \mathcal{D}^{p_{i+1}}{}_{[p_{i+1}}\cdots \mathcal{D}^{p_{a-1}}{}_{p_{a-1}]}  \nn
              &\qquad \qquad + (\sqrt x)^a\mathcal{D}^{p_1}{}_{[p_1}\cdots \mathcal{D}^{p_{a-1}}{}_{p_{a-1}]} \bigg]+\beta_{d+1}\sqrt{\frac{\xi}{\gamma}}  ,
\\
 \label{CBdef}
\mathcal{B}  =\;& \beta_0 + \sum_{a=1}^{d}\beta_a(\sqrt x)^a\mathcal{D}^{p_1}{}_{[p_1}\cdots \mathcal{D}^{p_a}{}_{p_a]}    .
\end{align}

After these decomposition, we see the only fields that contain time derivatives are $\gamma_{ij}$ and $\phi_A$. The conjugate momenta for the dynamical fields $\gamma_{ij}$ and $\phi_A$ are defined as
\begin{align} \label{piijcm}
\pi^{ij} &= \frac{\delta S}{\delta \dot{\gamma}_{ij}}  = \frac{\sqrt{\gamma}}{2}{\Omega} (K^{ij} - K \gamma^{ij})  - \frac{\sqrt{\gamma}}{2N} \gamma^{ij}\Omega^{,A} ( \dot{\phi}_A -N^k \pd_k \phi_A  ) ,\\
\label{piAcm}
 \pi^A   &= \frac{\delta S}{\delta \dot{\phi}_A} = - \sqrt{\gamma} K\Omega^{,A}
+\frac{\sqrt{\gamma}}{N} ( \dot{\phi}^A - N^i\pd_i \phi^A).
\end{align}
To pass to the Hamiltonian formalism, we need to solve these equations, and then replace $K_{ij}$ and $\dot{\phi}_A$ in favor of $\pi^{ij}$ and $\pi^A$. Note that these equations are linear in $K_{ij}$, $\dot{\phi}_A$, $\pi^{ij}$ and $\pi^A$, so we can straightforwardly solve these equations. But it is possible that primary constraints arise in this replacement. To see when this happens,  we define
\be \label{piiicm}
\pi\equiv \gamma_{ij}\pi^{ij} = \frac {1-d}{2} \Omega \sqrt{\gamma} K - \frac{d}{2} \Omega^{,A} \sqrt{\gamma} \bot^\mu \pd_\mu \phi_A   ,
\ee
and Eqs.~(\ref{piiicm}) and (\ref{piAcm}) can be combined to give
\be \label{eqpiiiA}
\left(
\begin{array}{c}
  \pi \\
  \pi^A
\end{array}
\right)
=
\left(
  \begin{array}{cc}
    \frac{1-d}{2}\Omega & -\frac d2\Omega_{,B} \\
    -\Omega^{,A} & \delta^A_B \\
  \end{array}
\right)
\left(
  \begin{array}{c}
    \sqrt{\gamma}K \\
    \sqrt{\gamma}\bot^\mu \pd_\mu \phi^B \\
  \end{array}
\right).
\ee
To replace $K_{ij}$ and $\phi_A$ in favor of $\pi^{ij}$ and $\pi^A$, we want to solve this matrix equation. Its invertibility depends on whether
\begin{align}\label{omegadef}
\omega &\equiv  -
\det
\left(
  \begin{array}{cc}
    \frac{1-d}{2}\Omega & -\frac d2\Omega_{,B} \\
    -\Omega^{,A} & \delta^A_B \\
  \end{array}
\right)
= \frac{d-1}{2}\Omega +\frac d2\Omega^{,A}\Omega_{,A}
\end{align}
is equal to zero or not.  This divides the general theory (\ref{startaction0}) into two sectors, which have to be dealt with separately. To simplify expressions in the following sections, we define
\be\label{thetadef}
\theta \equiv \frac{\Omega^{,A}\Omega_{,A}}{\Omega},
\ee
which leads to $\omega = (d-1+ d \theta)\Omega/2$. Note that for the minimally coupled case, we have $\Omega_{MC}=1, \omega_{MC}=({d-1})/{2}$ and $\theta_{MC}=0$.

\section{Hamiltonians and their constraint systems} \label{sec:consys}

In this section, we derive the Hamiltonian along with all possible constraints for the two sectors identified in the last section. Although arising via different ways, we will see that there are two second class constraints in either of the two sectors. In $d+1$ dimensions, after integrating out $n^i$, the spatial metric has $d(d+1)/2$ components and its phase space is $d(d+1)$ dimensional. With two constraints, the constrained phase space becomes $(d-1)(d+2)$ dimensional, which describes $(d-1)(d+2)/2$ physical degrees of freedom. This number of degrees of freedom matches the counting from the little group analysis, i.e., the symmetric two index tensor representation of SO($d$) is $(d-1)(d+2)/2$ dimensional. Thus, in any dimensions, as long as there are two constraints generated, the BD ghost is eliminated. However, we will also show that the sector $(d-1)\Omega= -d \Omega^{,A} \Omega_{,A}$ still suffers from ghost instabilities, except for the case of $d=1$.

\subsection{Sector $(d-1)\Omega\neq -d \Omega^{,A} \Omega_{,A}$}
\label{sec:absBDg}

For this sector,  we can solve Eq.~(\ref{eqpiiiA}) and get
\be
\left(
  \begin{array}{c}
    \sqrt{\gamma}K \\
    \sqrt{\gamma}\bot^\mu \pd_\mu \phi^A \\
  \end{array}
\right)
=\left(
  \begin{array}{cc}
    -\omega^{-1} & -\frac d2\omega^{-1}\Omega_{,B} \\
    -\omega^{-1}\Omega^{,A} &  \delta^A_B - \frac d2\omega^{-1}\Omega^{,A} \Omega_{,B} \\
  \end{array}
\right)
\left(
\begin{array}{c}
  \pi\\
  \pi^B
\end{array}
\right).
\ee
With the help of Eq.~(\ref{piijcm}), we can replace all the time derivatives with the corresponding conjugate momenta
\begin{align}
&\phantom{=}~\mathcal{L}_R + \mathcal{L}_S  \\
&=  N\sqrt{\gamma}\left[ \frac{\Omega}{2}{}^{(d)\!}R- D_iD^i\Omega
-\frac12 \pd_i \phi_A \pd^i \phi^A-V \right]  +\frac{N}{\sqrt{\gamma}} \bigg[  \frac{2}{\Omega} \pi^{ij}\pi_{ij}+\frac12 \pi^A \pi_A \nn
&~~ - \frac{1+\theta}{\omega}\pi^2 -  \frac{1}{\omega} \pi\Omega_{,A}\pi^A -
 \frac{d}{4\omega}(\Omega_{,A}\pi^A)^2 \bigg],
\end{align}
where $\omega$ and $\theta$ are defined in Eqs.~(\ref{omegadef}) and (\ref{thetadef}) respectively.  Also, we have
\begin{align}
\pi^{ij} \dot{\gamma}_{ij}
  &\dot{=}~\frac{N}{\sqrt{\gamma}} \left[\frac{4}{\Omega} \pi^{ij}\pi_{ij}
 - \frac{2+2\theta}{\omega}\pi^2 -  \frac{1}{\omega} \pi\Omega_{,A}\pi^A \right] -  2N_i D_j (\pi^{ij}),
 \\
 \pi^A \dot{\phi}_A &= \frac{N}{\sqrt{\gamma}} \left[
 \pi^A \pi_A -  \frac{1}{\omega} \pi \Omega_{,A}\pi^A -
 \frac{d}{2\omega} (\Omega_{,A}\pi^A)^2 \right]  +N^i \pi^A \pd_i\phi_A,
\end{align}
where $\dot{=}$ means ``equal up to some boundary terms''. Note that $\pi^{ij}$ is a $d$-D tensor density, whose $d$-D covariant derivative is defined as $D_j (\pi^{ij})=\sqrt{\gamma} D_j (\pi^{ij}/\sqrt{\gamma})$.

Now, the Hamiltonian is given by
\be \label{hamil}
H=\int \ud^d x \mathcal{H} =\int \ud^d x \left[ \pi^{ij} \dot{\gamma}_{ij}+ \pi^A \dot{\phi}_A -\mathcal{L} \right]=\int \ud^d x (\mathcal{H}_0  - N\mathcal{C}),
\ee
where
\begin{align}
\mathcal{H}_0 &= -(Ln^i+L^i)\mathcal{R}_i-L\sqrt{\gamma}\mathcal{A}  , \label{HH0general}\\
\mathcal{C}  &= \mathcal{R}+\mathcal{R}_i \mathcal{D}^i{}_jn^j+ \sqrt{\gamma}\mathcal{B} \label{Cgeneral},
\end{align}
and
\begin{align}
\label{Rexp}
\mathcal{R} &= \sqrt{\gamma}\left[ \frac{\Omega}{2}{}^{(d)\!}R- D_iD^i\Omega
-\frac12 \pd_i \phi_A \pd^i \phi^A-V \right]
 -\frac{1}{\sqrt{\gamma}} \bigg[  \frac{2}{\Omega} \pi^{ij}\pi_{ij}
+\frac12 \pi^A \pi_A \nn
&~~~ -  \frac{1+\theta}{\omega}\pi^2 -  \frac{1}{\omega} \pi \Omega_{,A}\pi^A -
 \frac{d}{4\omega} (\Omega_{,A}\pi^A)^2 \bigg],
 \\
 \label{Riexp}
\mathcal{R}_i &= 2\gamma_{ik}D_j (\pi^{kj}) -\pi^A \pd_i\phi_A.
\end{align}
$\mathcal{A}$ and $\mathcal{B}$ are defined in Eqs.~(\ref{CAdef}) and (\ref{CBdef}). Note that $\mathcal{H}_0$ and $\mathcal{C}$ contain the new shift vector $n^i$, which can in principle be integrated out by imposing their equations of motion $\pd \mathcal{L}/\pd n^i = -\pd \mathcal{H}/\pd n^i=0$. After some algebra, which we show in Appendix \ref{sec:Cieom}, this is equivalent to imposing the following conditions
\begin{align}\label{reducednieom}
\mathcal{C}_i  = \mathcal{R}_i - \sqrt\gamma{n^l\xi_{lj} \over \sqrt x}\sum_{a=1}^{d}\beta_a a(\sqrt x)^{a-1}\delta^j{}_{[i}\mathcal{D}^{p_1}{}_{p_1}\cdots \mathcal{D}^{p_{a-1}}{}_{p_{a-1}]}  =  0 .
\end{align}
From Eq.~(\ref{reducednieom}), we can in principle get an expansion of $n^i$ in terms of $\gamma_{ij},\pi^{ij},\phi_A,\pi^A$. We integrate out $n^i$ by substituting $n^i(\gamma_{ij},\pi^{ij},\phi_A,\pi^A)$ into the Hamiltonian (\ref{hamil}).

As $N$ is a Lagrange multiplier, it enforces a primary constraint
\be \label{prgs1}
\mathcal{C}(\gamma_{ij},\pi^{ij},\phi_A,\pi^A,n^i(\gamma_{ij},\pi^{ij},\phi_A,\pi^A)) \approx  0 ,
\ee
where we have used the weakly equality $\approx$, meaning an equality that holds on the submanifold of possible constraints. This is the only primary constraint in this sector.

Now, let us check whether there are secondary constraints by applying the Dirac-Bergman algorithm. This requires that the primary constraint (\ref{prgs1}) should be preserved under the time evolution, meaning $\ud \mathcal{C}/\ud t \approx 0$. On the other hand, the time evolution of $\mathcal{C}$ can be written as $\ud \mathcal{C}(x) / \ud t\approx\{\mathcal{C}(x), H\}$, where the Poisson bracket is defined as
\begin{align}
\{A(x), B(y) \} &= \int \ud^d z \( \frac{\delta A(x)}{\delta \gamma_{mn}(z)} \frac{\delta B(y)}{\delta \pi^{mn}(z)}
-\frac{\delta A(x)}{\delta \pi^{mn}(z)} \frac{\delta B(y)}{\delta \gamma_{mn}(z)} \) \nn
    &~~~+\int \ud^d z \( \frac{\delta A(x)}{\delta \phi_A(z)} \frac{\delta B(y)}{\delta \pi^A(z)}
-\frac{\delta A(x)}{\delta \pi^A(z)} \frac{\delta B(y)}{\delta \phi_A(z)} \)  .
\end{align}
So, for consistency, we require
\begin{align}\label{CH}
\int \ud^d y \left[\{\mathcal{C}(x),\mathcal{H}_0(y)\}- N(y)\{\mathcal{C}(x),\mathcal{C}(y)\}\right]\approx 0 .
\end{align}
 Now, if $\{\mathcal{C}(x),\mathcal{H}_0(y)\}\approx0$ and $\{\mathcal{C}(x),\mathcal{C}(y)\}\approx0$, no secondary constraint arises and the undetermined $N$ usually indicates a gauge degree. If $\{\mathcal{C}(x),\mathcal{C}(y)\}\not\approx 0$, we can determine $N$ via Eq.~(\ref{CH}) and no secondary constraint arises either. We will show that in the general sector a third case applies where $\{\mathcal{C}(x),\mathcal{C}(y)\}\approx0$ and $\{\mathcal{C}(x),\mathcal{H}_0(y)\}\not\approx 0$, and one obtains a secondary constraint given by $\{\mathcal{C}(x),\mathcal{H}_0(y)\} \approx 0$. If it was one of the first two cases, the Dirac-Bergman algorithm would be over. For the third case, which is the case we are facing, one has to further check the consistency of the generated secondary constraint, and so on, until the Lagrange multiplier $N$ is either determined (the second case) or can not be determined since the consistency requirements are trivially satisfied (the first case).

First, we show that this sector corresponds to the third case, that is, we show that $\{\mathcal{C}(x),\mathcal{C}(y)\}\approx0$ and $\{\mathcal{C}(x),\mathcal{H}_0(y)\}\not\approx 0$. To simplify our calculation, we note that $\delta\mathcal{C}/\delta n^k=\mathcal{C}_i \delta (\mathcal{D}^i{}_jn^j)/\delta n^k=0$, thanks to the $n^i$ equation of motion $\mathcal{C}_i=0$, which allows us to expand $\delta\mathcal{C}$ as
\begin{align}
\delta \mathcal{C} = \left.\frac{\pd \mathcal{C}}{\pd \gamma_{mn}}\right|_{n^i} \delta \gamma_{mn} + \left.\frac{\pd \mathcal{C}}{\pd \pi^{mn}}\right|_{n^i}  \delta \pi^{mn} +\left.\frac{\pd \mathcal{C}}{\pd \phi_A}\right|_{n^i} \delta \phi_A + \left.\frac{\pd \mathcal{C}}{\pd \pi^A}\right|_{n^i}  \delta \pi^A    .
\end{align}
Also, since $\{ \mathcal{C}(x), \mathcal{C}(y) \}$ is anti-symmetric in terms of exchanging $x$ and $y$, we can drop any terms that are symmetric under this exchange, notably terms proportional to $\delta^d(x-y)$. Then, expanding $\mathcal{C}$ using Eq.~(\ref{Cgeneral}), we get
\begin{align}\label{CC}
\{ \mathcal{C}(x), \mathcal{C}(y) \} & = \{\mathcal{R}(x), \mathcal{R}(y)\} + \{\mathcal{R}_i(x), \mathcal{R}_j(y)\} \mathcal{D}^i{}_k n^k(x) \mathcal{D}^j{}_ln^l(y) \nn
  &~~~+\left[ \{\mathcal{R}(x), \mathcal{R}_i(y)\} \mathcal{D}^i{}_k n^k(y)
  + S^{mn}(x)\frac{\delta \mathcal{R}_i(y)}{\delta \pi^{mn}(x)} \mathcal{D}^i{}_kn^k(y)  - (x\leftrightarrow y)\right]  ,
\end{align}
where
\begin{align} \label{smn}
S^{mn}(x) = \mathcal{R}_j\frac{\pd (\mathcal{D}^j{}_r n^r)}{\pd \gamma_{mn}}(x)+\frac{\pd (\sqrt{\gamma}\mathcal{B})}{\pd \gamma_{mn}}(x)  .
\end{align}

After a lengthy calculation, which is shown in Appendix \ref{sec:genpoissonb}, we can get that the Poisson brackets of $\mathcal{R}$ and $\mathcal{R}_i$ in this sector satisfy the following algebra
\begin{align}
\label{RR}
\{\mathcal{R}(x), \mathcal{R}(y)\}  &=-\left[\mathcal{R}^i(x)\pd_{x^i} \delta^d(x-y) - (x\leftrightarrow y)\right]  ,\\
\label{RRi}
\{\mathcal{R}(x), \mathcal{R}_i(y)\}&=-\mathcal{R}(y)\pd_{x^i} \delta^d(x-y)  ,  \\
\label{RiRi}
\{\mathcal{R}_i(x), \mathcal{R}_j(y)\} &= -\[\mathcal{R}_j(x)\pd_{x^i} \delta^d(x-y) - \mathcal{R}_i(y)\pd_{y^j} \delta^d(x-y)\]  .
\end{align}
To further facilitate the evaluation of these Poisson brackets, we introduce time-independent smoothing functions $f(x)$ and $g(y)$ that decay quickly at infinities and define
\begin{align}
F \equiv \int\ud^dx~f(x)\mathcal{C}(x), \quad G \equiv \int\ud^dy~g(y)\mathcal{C}(y) .
\end{align}
We then calculate the smeared Poisson bracket
\begin{align}
\{F , G\}=\int\ud^dx\ud^dy~f(x)g(y)\{\mathcal{C}(x),\mathcal{C}(y)\} .
\end{align}
and extract $\{\mathcal{C}(x), \mathcal{C}(y)\}$ at the last step.

Using Eqs.~(\ref{RR}-\ref{RiRi}), we can simplify the smeared Poisson bracket as
\be \label{fgp}
\{F,G\} = -\int d^dx \Big(f \, \pd_i g-g\, \pd_i f\Big)\, \mathcal{P}^i\,,
\ee
where
\be
\mathcal{P}^i=(\mathcal{R} + \mathcal{R}_j \mathcal{D}^j_{~k}n^k) \mathcal{D}^i_{~l}n^l +\mathcal{R}^i+  2S^{il}\gamma_{lj}
\mathcal{D}^j_{~k}n^k \, .
\ee
Further simplification of $\mathcal{P}^i$ requires simplifying $S^{mn}$:
\begin{align}
S^{mn}={1\over2}\sqrt\gamma\gamma^{nm}\mathcal{B} + \mathcal{R}_i\frac{\pd(\mathcal{D}^i{}_jn^j)}{\pd\gamma_{mn}} + \sqrt\gamma\frac{\pd\mathcal{B}}{\pd\gamma_{mn}} .
\end{align}
From Eq.~(\ref{Ddefrel}), we can derive
\begin{align}
\frac{\pd}{\pd\gamma_{mn}}\text{tr}[(\sqrt x \mathcal{D})^a] = \frac a2 (\sqrt x)^{a-2}\xi_{lk}(\mathcal{D}^{a-2})^k{}_i\frac{\pd\gamma^{il}}{\pd\gamma_{mn}} - a(\sqrt x)^{a-2}n^i\xi_{ij}(\mathcal{D}^{a-1})^j{}_k\frac{\pd(\mathcal{D}^k{}_ln^l)}{\pd\gamma_{mn}} ,
\end{align}
making use of which we get
\begin{align}
\frac{\pd\mathcal{B}}{\pd\gamma_{mn}} =& -{1\over2}\gamma^{mi} \sum_{a=1}^{d} a\beta_a (\sqrt x)^{a-2} \xi_{lk}(\mathcal{D}^{-1})^k{}_j \delta^j{}_{[i}\mathcal{D}^{p_1}{}_{p_1} \cdots \mathcal{D}^{p_{a-1}}{}_{p_{a-1}]} \,\gamma^{ln} \nn
& - \sum_{a=1}^{d} \beta_a a(\sqrt x)^{a-1}{n^k\xi_{kj} \over \sqrt x} \delta^j{}_{[i}\mathcal{D}^{p_1}{}_{p_1} \cdots \mathcal{D}^{p_{a-1}}{}_{p_{a-1}]} \frac{\pd(\mathcal{D}^i{}_ln^l)}{\pd\gamma_{mn}} \,.
\end{align}
Then, using the $n^i$ equation of motion $\mathcal{C}_i=0$, one gets
\begin{align}\label{Smn}
S^{mn} =  {1\over2}\sqrt\gamma\( \mathcal{B}\gamma^{mn}-\bar{\mathcal{B}}^{mn} \) ,
\end{align}
where
\begin{align}\label{bmn}
\bar{\mathcal{B}}^{mn}\equiv &\gamma^{mi} \sum_{a=1}^{d} a\beta_a (\sqrt x)^{a-2} \xi_{lk}(\mathcal{D}^{-1})^k{}_j \delta^j{}_{[i}\mathcal{D}^{p_1}{}_{p_1} \cdots \mathcal{D}^{p_{a-1}}{}_{p_{a-1}]}\,\gamma^{ln}\,.
\end{align}
Therefore, we have
\be
\mathcal{P}^i  = \mathcal{C} \mathcal{D}^i{}_l n^l  \approx 0 .
\ee
and consequently $\{ \mathcal{C}(x), \mathcal{C}(y) \}\approx 0$.

We show $\{ \mathcal{C}(x), \mathcal{H}_0(y) \}\not\approx 0$ as follows. Similar to the calculation of $\{\mathcal{C}(x), \mathcal{C}(y)\}$, we expand $\delta\mathcal{C}$ and $\delta\mathcal{H}_0$ at fixed $n^i$, and we have
\begin{align}
\!\{{\cal C}(x) , {\cal H}_0(y)\}&
\!=\!-\{\mathcal{R}(x),\mathcal{R}_i(y)\}(Ln^i\!+\!L^i)(y)-D^i_{~k}n^k(x)\,
\{\mathcal{R}_i(x),\mathcal{R}_j(y)\}(Ln^j\!+\!L^j)(y) \nn
&~~+L\frac{\delta \mathcal{R}(x)}{\delta\pi^{mn}(y)}\,\sqrt{\gamma}A^{mn}(y)
+ L\,D^i_{~k}n^k(x)\,\frac{\delta \mathcal{R}_i(x)}{\delta\pi^{mn}(y)}
\sqrt{\gamma}A^{mn}(y)\nn
&~~-S^{mn}(x)\,\frac{\delta \mathcal{R}_i(y)}{\delta\pi^{mn}(x)}\,(Ln^i\!+\!L^i)(y)
- \sqrt{\gamma} \frac{\pd \mathcal{B}}{\pd \phi_A}\frac{\pd\mathcal{R}_i}{\pd\pi^A} (Ln^i+L^i)\delta^d(x-y) \nn
&~~+L\sqrt{\gamma} \frac{\pd \mathcal{A}}{\pd \phi_A}\frac{\pd\mathcal{R}}{\pd\pi^A} \delta^d(x-y) +L \sqrt{\gamma} \frac{\pd \mathcal{A}}{\pd \phi_A}\frac{\pd\mathcal{R}_i}{\pd\pi^A}D^i{}_kn^k \delta^d(x-y),
\end{align}
where $S^{mn}$ is given by Eq.~(\ref{Smn}) and
\begin{align}
A^{mn} \equiv {1\over\sqrt\gamma} \frac{\pd(\sqrt\gamma\mathcal{A})}{\pd\gamma_{mn}} \,.
\end{align}
After some algebra, we have
\begin{align}\label{C2}
\!\{{\cal C}(x) , {\cal H}_0(y)\} & ={\cal C}  D_i(Ln^i+L^i) +
 L A^{mn}\[-{4\over\Omega}\pi_{mn}+{2+2\theta \over \omega}\pi\gamma_{mn}+{1 \over \omega}\Omega_A\pi^A \gamma_{mn} \] \nn
&~~+2L\sqrt{\gamma} D_mA^{mn} \gamma_{ni} \mathcal{D}^i_{~k}n^k + \(\mathcal{R}_j \mathcal{D}^i_{~k}n^k - \sqrt\gamma\gamma_{jk}\bar{\mathcal{B}}^{ki}
\) D_i(Ln^j+L^j) \nn
&~~+(D_i \mathcal{R} + \mathcal{D}^j_{~k}n^k D_i\mathcal{R}_j)(Ln^i+L^i) + \sqrt\gamma \frac{\pd\mathcal{B}}{\pd\phi_A} \pd_i\phi_A (Ln^i+L^i) \nn
 &~~ + L \frac{\pd\mathcal{A}}{\pd\phi_A} \( -\pi_A + {1 \over \omega}\Omega_A\pi + {d \over 2\omega}\Omega_B\pi^B \Omega_A - \sqrt\gamma \pd_i\phi_A \mathcal{D}^i_{~k}n^k \),
\end{align}
where $\bar{\mathcal{B}}^{mn}$ is defined by Eq.~(\ref{bmn}). This is generally not zero after using $\mathcal{C}\approx 0$. So we get a secondary constraint:
\be
\mathcal{C}^{(2)}(x)\equiv \int\ud^dy~\{ \mathcal{C}(x), \mathcal{H}_0(y) \} \approx 0.
\ee

As mentioned above, we need to check whether $\mathcal{C}^{(2)}$ is preserved in time $\ud \mathcal{C}^{(2)}(x)/\ud t\approx 0$. That is, one has to check the following equation
\begin{align}
  \{\mathcal{C}^{(2)}(x),H\}
= \int \ud^d y \left[\{\mathcal{C}^{(2)}(x),\mathcal{H}_0(y)\}- N(y)\{\mathcal{C}^{(2)}(x),\mathcal{C}(y)\} \right] \approx 0.
\end{align}
 $\{\mathcal{C}^{(2)}(x),\mathcal{C}(y)\}$ generally does not vanish weakly, as it does not vanish weakly even for the dGRT case \cite{Hassan:2011ea}. So this consistency requirement fixes $N$, which terminates the Dirac-Bergman algorithm. Therefore, we have obtained all the constraints in this sector, which are the second class constraints $\mathcal{C}$ and $\mathcal{C}^{(2)}$.

\subsection{Sector $(d-1)\Omega= -d \Omega^{,A}\Omega_{,A}$}

In this sector, the matrix equation~(\ref{eqpiiiA}) can not be inverted, as the $\mathcal{N}+1$ sub-equations are related. As the rank of the matrix in Eq.~(\ref{eqpiiiA}) now is $\mathcal{N}$, there is one primary constraint relating $\gamma_{ij},\pi^{ij},\phi_A,\pi^A$. Making use of the relation $(d-1)\Omega= -d \Omega^{,A}\Omega_{,A}$, we see the primary constraint is
\be\label{Clambdadef}
\mathcal{C}_\lambda=\pi+{d\over2}\Omega_{,A}\pi^A\approx 0 \,.
\ee
With the help of this constraint, we can now replace $K_{ij}$ and $\dot{\phi_A}$ in favor of $\pi_{ij}$ and $\pi_A$:
\begin{align}
&\phantom{=}~\mathcal{L}_R + \mathcal{L}_S  \nn
&=  N\sqrt{\gamma}\left[ \frac{\Omega}{2}{}^{(d)\!}R- D_iD^i\Omega
-\frac12 \pd_i \phi_A \pd^i \phi^A-V \right]  + \frac{N}{\sqrt{\gamma}} \bigg[  \frac{2}{\Omega} \pi^{ij}\pi_{ij} + \frac12 \pi^A\pi_A - \frac{2}{\Omega  d}\pi^2 \bigg]  ,
\end{align}
and  the Hamiltonian in this sector is given by
\be \label{hamil2}
H=\int \ud^d x \mathcal{H} =\int \ud^d x \left[ \pi^{ij} \dot{\gamma}_{ij}+ \pi^A \dot{\phi_A} -\mathcal{L} \right]=\int \ud^d x (\mathcal{H}_0  - N\mathcal{C}),
\ee
where
\begin{align}
\mathcal{H}_0 &= -(Ln^i+L^i)\mathcal{R}_i-L\sqrt{\gamma}\mathcal{A}  , \label{HH0}\\
\mathcal{C}  &= \mathcal{R}+\mathcal{R}_i \mathcal{D}^i{}_jn^j+ \sqrt{\gamma}\mathcal{B} \label{C}.
\end{align}
Here
\begin{align}\label{Rprime}
\mathcal{R} &= \sqrt{\gamma}\left[ \frac{\Omega}{2}{}^{(d)\!}R- D_iD^i\Omega
-\frac12 \pd_i \phi_A \pd^i \phi^A-V \right] -\frac{1}{\sqrt{\gamma}} \[  \frac{2}{\Omega} \pi^{ij}\pi_{ij} + \frac12 \pi^A\pi_A
 - \frac{2}{\Omega d} \pi^2 \] , \\
 \label{Ridef2}
\mathcal{R}_i &= 2\gamma_{ik}D_j (\pi^{kj}) -\pi^A \pd_i\phi_A \,.
\end{align}
$\mathcal{A}$ and $\mathcal{B}$ are defined in Eqs.~(\ref{CAdef}) and (\ref{CBdef}). This only differs from the general sector Hamiltonian (\ref{hamil}) in $\mathcal{R}$. Interestingly, we can formally obtain this Hamiltonian, by taking the limit $\theta \rightarrow \infty$ of the general sector Hamiltonian (\ref{hamil}). Similar to the general sector case,  we can in principle integrate out the new shift vector $n^i$ by imposing the following conditions
\begin{align}
\mathcal{C}_i  = \mathcal{R}_i - \sqrt\gamma{n^l\xi_{lj} \over \sqrt x}\sum_{a=1}^{d}\beta_a a(\sqrt x)^{a-1}\delta^j{}_{[i}\mathcal{D}^{p_1}{}_{p_1}\cdots \mathcal{D}^{p_{a-1}}{}_{p_{a-1}]}  =  0 .
\end{align}
Substituting $n^i(\gamma_{ij},\pi^{ij},\phi_A,\pi^A)$ into the Hamiltonian (\ref{hamil2}), the  equation of motion for $N$ gives rise to another primary constraint
\be
\mathcal{C}(\gamma_{ij},\pi^{ij},\phi_A,\pi^A,n^i(\gamma_{ij},\pi^{ij},\phi_A,\pi^A)) =  0 .
\ee

Since there is a hidden primary constraint arising from switching from the ``velocities'' to conjugate momenta, we can define the total Hamiltonian by adding $\mathcal{C}_\lambda$ to the Hamiltonian
\begin{align}
 H_T = H+\int \ud^d x \lambda\mathcal{C}_\lambda = \int \ud^d x (\mathcal{H}_0  - N\mathcal{C}+ \lambda\mathcal{C}_\lambda),
\end{align}
where $\lambda$ is a Lagrangian multiplier. Therefore, in this sector  there are two primary constraints $\mathcal{C}$ and $\mathcal{C}_\lambda$.

Now, we apply the Dirac-Bergman algorithm to get potential secondary constraints. To this end, we require the consistency conditions $\ud \mathcal{C}(x)/\ud t \approx \{\mathcal{C}(x),H_T\}\approx 0$ and  $\ud\mathcal{C}_\lambda(x) \ud t\approx \{\mathcal{C}_\lambda(x),H_T\}\approx 0$, i.e.,
\begin{align}
\label{sccc1}
 \int \ud^d y \left[\{\mathcal{C}(x),\mathcal{H}_0(y)\}- N(y)\{\mathcal{C}(x),\mathcal{C}(y)\} + \lambda(y)\{\mathcal{C}(x),\mathcal{C}_\lambda(y)\}\right]&\approx 0 ,\\
\label{sccc2}
 \int \ud^d y \left[\{\mathcal{C}_\lambda(x),\mathcal{H}_0(y)\}- N(y)\{\mathcal{C}_\lambda(x),\mathcal{C}(y)\} + \lambda(y)\{\mathcal{C}_\lambda(x),\mathcal{C}_\lambda(y)\}\right]&\approx 0.
\end{align}

To evaluate these Poisson brackets, we can make use of the following Poisson brackets for $\mathcal{R}$, $\mathcal{R}_i$ and $\mathcal{C}_\lambda$:
\begin{align}
\{\mathcal{R}(x), \mathcal{R}_i(y)\}&=-\mathcal{R}(y)\pd_{x^i} \delta^d(x-y) ,
\\
\{\mathcal{R}_i(x), \mathcal{R}_j(y)\} &= -\[\mathcal{R}_j(x)\pd_{x^i} \delta^d(x-y) - \mathcal{R}_i(y)\pd_{y^j} \delta^d(x-y)\],
\\
\label{scca3}
\{\mathcal{R}_i(x), \mathcal{C}_\lambda(y)\}&=-\mathcal{C}_\lambda(x)\pd_{x^i}\delta^d(x-y) ,
\\
\{\mathcal{C}_\lambda(x), \mathcal{C}_\lambda(y)\}&=0  ,
\\
\{\mathcal{R}(x),\mathcal{R}(y)\} &=-\[(\mathcal{R}^i-\frac2d D^i\mathcal{C}_{\lambda})(x)\pd_{x^i} \delta^d(x-y) - (x\leftrightarrow y)\],
\\
\label{scca6}
\{\mathcal{R}(x),\mathcal{C}_\lambda(y)\} =& \bigg[ -{1\over2}\mathcal{R} - \sqrt\gamma{d\over2}D_i\Omega_AD^i\phi^A - \sqrt\gamma{d-1 \over 4}\pd_i\phi_A\pd^i\phi^A - \sqrt\gamma{d+1 \over 2}V \nn
& - \sqrt\gamma{d\over2}V^A\Omega_A + {d \over 4\sqrt\gamma}{(\Omega_A\pi^A)^2\over\Omega} + {d-1 \over 4\sqrt\gamma}\pi^A\pi_A \bigg](x)\delta^d(x-y) ,
\end{align}
which are calculated in Appendix \ref{sec:appsc}.  Following steps similar to the general sector case, we can get
\begin{align}
\int\ud^dx\ud^dy~f(x)g(y)\{\mathcal{C}(x),\mathcal{C}(y)\} = -\int d^dx \Big(f \, \pd_i g-g\, \pd_i f\Big)\, \mathcal{P}^i,
\end{align}
where $f(x)$ and $g(y)$ are smoothing functions and $\mathcal{P}^i$ can be simplified to
\be
\mathcal{P}^i  = \mathcal{C} \mathcal{D}^i{}_l n^l  + \frac2d\frac{\mathcal{C}_{\lambda}}{\sqrt\gamma}\pd^i\sqrt\gamma -\frac2d\pd^i\mathcal{C}_{\lambda} .
\ee
So, despite the differences from the general sector, $\mathcal{P}^i$ again vanishes weakly and we have $\{\mathcal{C}(x),\mathcal{C}(y)\}\approx0$ in this sector.

We also have to calculate $\{\mathcal{C}(x),\mathcal{C}_\lambda(y)\}$. Since $\pd\mathcal{C}_\lambda/\pd n^i=0$ and $\pd\mathcal{C}/\pd n^i=0$, we can expand $\{\mathcal{C}(x),\mathcal{C}_\lambda(y)\}$ at fixed $n^i$
\begin{align}\label{CCnew}
\{\mathcal{C}(x),\mathcal{C}_\lambda(y)\} = \{\mathcal{R}(x),\mathcal{C}_\lambda(y)\} + \mathcal{D}^i{}_jn^j(x)\{\mathcal{R}_i(x),\mathcal{C}_\lambda(y)\} + S^{mn}(x){\delta\mathcal{C}_\lambda(y) \over \delta\pi^{mn}(x)},
\end{align}
where $S^{mn}$ is given by Eq.~(\ref{Smn}). Using Eqs.~(\ref{scca3}) and (\ref{scca6}), we then get
\begin{align}
\{\mathcal{C}(x),\mathcal{C}_\lambda(y)\} =& \bigg[ -{1\over2}\mathcal{R} - \sqrt\gamma{d\over2}D_i\Omega_AD^i\phi^A - \sqrt\gamma{d-1 \over 4}\pd_i\phi_A\pd^i\phi^A - \sqrt\gamma{d+1 \over 2}V \nn
& - \sqrt\gamma{d\over2}V^A\Omega_A + {d \over 4\sqrt\gamma}{(\Omega_A\pi^A)^2\over\Omega} + {d-1 \over 4\sqrt\gamma}\pi^A\pi_A \bigg](x)\delta^d(x-y) \nn
& + S^{mn}(x)\gamma_{mn}(y)\delta^d(x-y).
\end{align}
Therefore, due to the graviton potential terms, we have $\{\mathcal{C}(x),\mathcal{C}_\lambda(y)\}\not\approx 0$.

Although not essential for completing the Dirac-Bergman algorithm, we also compute $\{\mathcal{C}(x),\mathcal{H}_0(y)\}$ and $\{\mathcal{C}_\lambda(x),\mathcal{H}_0(y)\}$ for the completeness of the Hamiltonian formulation. By explicit calculation, we get that $\{\mathcal{C}(x),\mathcal{H}_0(y)\}$ is given by the expression (\ref{C2}) with $\theta$ taken to $\infty$, which does not vanish weakly.  Also,
\begin{align}
\{\mathcal{C}_\lambda(x),\mathcal{H}_0(y)\}
& \approx \( -\gamma_{mn}\sqrt\gamma A^{mn} - {d\over2}\Omega_{,A}\sqrt\gamma\frac{\pd\mathcal{A}}{\pd\phi_A} \)\delta^d(x-y) \not\approx 0\,.
\end{align}

Now, with all the Poisson brackets in Eqs.~(\ref{sccc1}) and (\ref{sccc2}) calculated, we see that $\lambda$ can be determined from  Eq.~(\ref{sccc1}) and then $N$ can be determined from Eq.~(\ref{sccc2}). Therefore, the Dirac-Bergman algorithm terminates here and there are no secondary constraints generated, and the constraints in this sector are the second class constraints $\mathcal{C}$ and $\mathcal{C}_\lambda$. Therefore there is no BD ghost in this sector either.

However, this does not mean this special sector is free from problems. Indeed, as we shall see this sector suffers from ghost instabilities, except for the case of $d=1$. From the relation $(d-1)\Omega = -d\Omega^{,A}\Omega_{,A}$, we have $\Omega<0$, except for the case of $d=1$. When $\Omega<0$, the graviton kinetic term has the wrong sign. But we can't exclude all the cases with $\omega=0$ and $d>1$ based on this yet. One should keep in mind that the overall sign in the action does not affect the equations of motion. If all the kinetic terms in the action, after diagonalization, have the wrong sign, this ghost can be ``eliminated'' by redefining the action with an overall sign. This turns out to be the case if there is only one scalar.

To see this, we first note that for the case of one scalar we have
\be
\Omega(\phi) = -\frac{d-1}{4d}(\phi+c)^2, \quad c=const..
\ee
Now, conformally transforming to $\tilde{g}_{\mu\nu} = (-\Omega)^{\frac{2}{d-1}} g_{\mu\nu}$ and neglecting some boundary terms, we get the redefined action is given by
\be \label{Spconfm}
{S}' = -S =  \int\ud^{d+1}x \sqrt{-\tilde{g}} \left[ \frac{\tilde{R}}{2} + \sum_{a=1}^{d+1}\tilde{\alpha}_a(\phi)e_a(\tilde{\mathcal{{K}}})  - \tilde{V}(\phi) \right] .
\ee
where $\tilde{\mathcal{K}}^\mu_\nu=\delta^\mu_\nu - \sqrt{\tilde{g}^{\mu\rho}f_{\rho\nu}}$, $ \tilde{V}(\phi)= -(-\Omega)^{-\frac{d+1}{d-1}}{V}(\phi)+\tilde{\alpha}_0(\phi)$ and  $\tilde{\alpha}_n(\phi)$ and $\beta_n(\phi)$ are related by
\begin{align}
\tilde{\alpha}_n = (-1)^{n+1} \sum_{a=n}^{d+1} \frac{(d+1-n)!}{(d+1-a)!(a-n)!}  (-\Omega)^{\frac{a}{d-1}-\frac{d+1}{d-1}} \beta_a .
\end{align}
For the case of multiple scalars, however, this conformal transformation trick does not work, as it can at most ``absorb'' one scalar kinetic term. Therefore, the case of multiple scalars suffers from ghost instabilities. 

However, the case of (\ref{Spconfm}) still suffers from a hidden ghost instability. In the action (\ref{Spconfm}), $\phi$ is an auxiliary field and we can integrate it out, which gives rise to a massive gravity action for the metric $\tilde{g}_{\mu\nu}$. Now, from our constraint system analysis above, we know that this massive gravity for $\tilde{g}_{\mu\nu}$ has 6 degrees of freedom, which means there is a ``second BD ghost'' in this case. Therefore, all the cases with $\omega=0$ and $d>1$ suffer from ghost instabilities.

For the case with $\omega=0$ and $d=1$, we have $\Omega=const.$, which can be positive or negative. Note that, for this case,  even $\Omega<0$ does not necessarily imply there is a ghost, because in 1+1 dimensions there is only one component for the spatial metric $\gamma_{ij}=\gamma_{11}$. As we have shown, this only potential degree of freedom is eliminated by two constraints, therefore the massive graviton does not have any degree of freedom, and the scalars have the right kinetic terms.

\section{Discussions}
\label{sec:conclu}

In this paper, we have generalized mass-varying massive gravity \cite{Huang:2012pe} to a general scalar massive-tensor theory. Mass-varying massive gravity has been shown to give rise to stable fully FRW solutions if a few conditions are satisfied \cite{Gumrukcuoglu:2013nza}. It is, however, not clear how restricting these conditions are. In any event, stability of the FRW solutions is only a first-step requirement and extending its possible theory space may be useful to obtain an eventually phenomenologically viable model, in which the functions $\alpha_i(\phi_A)$ and $V(\phi_A)$ are to be chosen appropriately.

We have allowed for generic mixing between the kinetic terms and the mass terms of the massive graviton and the scalars. One may alway de-mix the kinetic mixing between the graviton and scalars by a conformal transformation, but matter may couple to the graviton in the Jordan frame. Also, the mass matrix mixing has not been considered in previous generalizations of the dRGT model and in the studies of their background FRW solutions and perturbations. As in the Stuckelberg language this mass matrix mixing can give rise to kinetic terms for the Stuckelberg fields, it is therefore interesting to see how the perturbative analysis of the FRW solution will change in presence of this mass mixing, as well as the kinetic mixing, which we leave for further work.

We have derived the Hamiltonian formulation for our generalized massive gravity theory. We see that, depending on the kinetic coupling $\Omega(\phi_A)$,  two sectors arise when passing to the Hamiltonian formulation. The Hamiltonian and its constraints for the two sectors have been derived separately. In the general sector there is one primary and one secondary constraint, while there are two primary constraints and no secondary constraint in the special sector. Thus, the BD ghost is absent in both the two sectors. However, the special sector suffers from  ghost instabilities except for the $d=1$ case. We emphasize that, although most of the ghost instabilities in the special sector are obviously due to the fact that  $\Omega(\phi_A)<0$, it is necessary to calculate the constraint system for this sector to show that there is a ``second BD ghost'' for the special sector with one scalar and that the $d=1$ case is free from any obvious ghost instability.

We have formulated our model in arbitrary dimensions. As the dRGT graviton potential is in a sense very much unique by construction, we expect that there are also applications of dRGT-like massive gravity in a dimension other than four. For example, massive gravitons generally arise after the Kaluza-Klein compactification of a higher dimensional gravity theory. We expect the developed Hamiltonian formulation will be useful for computing energies of gravitational solutions in a large class of dRGT-like massive gravity, for numerically evolving the field configurations and even for discussing possible canonical quantization of this class of models. Finally, we note that low dimensional gravity theories sometimes are easy to solve, which may provide valuable lessons for quantizing gravity theories.

~\\
{\bf Acknowledgments}\\
We would like to thank for Loriano Bonora, A.~E.~Gumrukcuoglu, Chunshan Lin, Yun-Song Piao and Thomas Sotiriou for helpful discussions. QGH is supported by the project of Knowledge Innovation Program of Chinese Academy of Science and a grant from NSFC (grant NO. 10821504). SYZ acknowledges partial financial support from the European Research Council under the European Union's Seventh Framework Programme (FP7/2007-2013) / ERC Grant Agreement n.~306425 ``Challenging General Relativity'' and from the Marie Curie Career Integration Grant LIMITSOFGR-2011-TPS Grant Agreement n.~303537.

\appendix

\section{Reducing the $n^i$ equation of motion}  \label{sec:Cieom}

Here we simplify the $n^i$ equation of motion, which is
\begin{align} \label{motion}
\frac{\pd\mathcal{L}}{\pd n^k} &= -\frac{\pd\mathcal{H}_0}{\pd n^k}+N\frac{\pd\mathcal{C}}{\pd n^k}
\nn
&=L \mathcal{R}_k + L\sqrt\gamma\frac{\pd\mathcal{A}}{\pd n^k} +N \mathcal{R}_i\frac{\pd(\mathcal{D}^i{}_jn^j)}{\pd n^k} + N\sqrt\gamma\frac{\pd\mathcal{B}}{\pd n^k} = 0.
\end{align}
From Eq.~(\ref{Ddefrel}), a useful relation can be derived
\be\label{me1}
\frac{\pd}{\pd n^l}\text{tr}[(\sqrt x\mathcal{D})^a]=-a(\sqrt x)^{a-2}n^k\xi_{kj}(\mathcal{D}^{a-1})^j{}_i\frac{\pd (\mathcal{D}^i{}_tn^t)}{\pd n^l} ,
\ee
with which we can show that
\begin{align}
  \frac{\pd\mathcal{B}}{\pd n^k} =  -\sum_{a=1}^{d}\beta_a{n^r\xi_{rj} \over \sqrt x}\frac{\pd(\mathcal{D}^i{}_tn^t)}{\pd n^k} a (\sqrt x)^{a-1}\delta^j{}_{[i}\mathcal{D}^{p_1}{}_{p_1}\cdots \mathcal{D}^{p_{a-1}}{}_{p_{a-1}]}.
\end{align}
This leads to
\begin{align}
\frac{\pd\mathcal{C}}{\pd n^k} = \mathcal{C}_i\frac{\pd(\mathcal{D}^i{}_jn^j)}{\pd n^k},
\end{align}
where
\begin{align}
\mathcal{C}_i = \mathcal{R}_i - \sqrt\gamma{n^l\xi_{lj} \over \sqrt x}\sum_{a=1}^{d}\beta_a a(\sqrt x)^{a-1}\delta^j{}_{[i}\mathcal{D}^{p_1}{}_{p_1}\cdots \mathcal{D}^{p_{a-1}}{}_{p_{a-1}]}.
\end{align}

Similarly, we can show
\begin{align}\label{an}
\frac{\pd\mathcal{A}}{\pd n^k}
=& \sum_{a=1}^{d}\beta_a \bigg[-\sum_{i=1}^{a-1}(-1)^i\xi_{kj}[(\sqrt x\mathcal{D})^i]^j{}_{p_i}n^{p_i}\sqrt x \mathcal{D}^{p_{i+1}}{}_{[p_{i+1}}\cdots \sqrt x\mathcal{D}^{p_{a-1}}{}_{p_{a-1}]} \nn
 &  -{n^l\xi_{lk}\over\sqrt x}(\sqrt x)^{a-1}\mathcal{D}^{p_1}{}_{[p_1}\cdots \mathcal{D}^{p_{a-1}}{}_{p_{a-1}]} + T_1+T_2+T_3+T_4\bigg],
\end{align}
where
\begin{align}
T_1=&-\sum_{i=2}^{a-1}(-1)^in^l\xi_{lp_0}\bigg[ \sum_{m=1}^{i-1}\sqrt x\mathcal{D}^{p_0}{}_{p_1}\sqrt x\mathcal{D}^{p_1}{}_{p_2}\cdots\( \frac{\pd(\sqrt x\mathcal{D}^{p_{m-1}}{}_{p_m})}{\pd n^k} \)\sqrt x\mathcal{D}^{p_m}{}_{p_{m+1}} \nn
 & \cdots\sqrt x\mathcal{D}^{p_{i-2}}{}_{p_{i-1}} \bigg]\mathcal{D}^{p_{i-1}}{}_{p_i}n^{p_i}\sqrt x \mathcal{D}^{p_{i+1}}{}_{[p_{i+1}}\cdots \sqrt x\mathcal{D}^{p_{a-1}}{}_{p_{a-1}]} ,\nn
T_2=&\sum_{i=1}^{a-1}(-1)^in^l\xi_{lp_0}[(\sqrt x\mathcal{D})^{i-1}]^{p_0}_{p_{i-1}}\mathcal{D}^{p_{i-1}}{}_{p_i}n^{p_i}{n^c\xi_{cp_{i+1}}\over\sqrt x} \frac{\pd (\mathcal{D}^j{}_tn^t)}{\pd n^k} \nn
 & (a-i-1)(\sqrt x)^{a-i-2}\delta^{p_{i+1}}{}_{[j}\mathcal{D}^{p_{i+2}}{}_{p_{i+2}}\cdots \mathcal{D}^{p_{a-1}}{}_{p_{a-1}]} ,\nn
T_3=&-\sum_{i=1}^{a-1}(-1)^in^l\xi_{lp_0}[(\sqrt x\mathcal{D})^{i-1}]^{p_0}{}_{p_{i-1}} \frac{\pd (\mathcal{D}^{p_{i-1}}{}_{p_i}n^{p_i})}{\pd n^k}\sqrt x \mathcal{D}^{p_{i+1}}{}_{[p_{i+1}}\cdots \sqrt x\mathcal{D}^{p_{a-1}}{}_{p_{a-1}]} ,\nn
T_4=&-n^l\xi_{lj}\frac{\pd (\mathcal{D}^i{}_tn^t)}{\pd n^k}(a-1)(\sqrt x)^{a-2}\delta^j{}_{[i}\mathcal{D}^{p_1}{}_{p_1}\cdots \mathcal{D}^{p_{a-2}}{}_{p_{a-2}]}.
\end{align}
To further simplify $\pd\mathcal{A}/\pd n^k$, we note another useful relation that can come out of the properties of $\mathcal{D}^i{}_j$:
\begin{align}\label{me2}
n^T\xi\bigg[ \sqrt x \mathcal{D}^{i-1}+\frac{1}{\sqrt x} & \mathcal{D}^{m-1}n^T\xi\mathcal{D}^{i-m}n \bigg]\frac{\pd(\mathcal{D}n)}{\pd n^k} = n^T\xi\[ \sqrt x \mathcal{D}^i+\frac{1}{\sqrt x}I(n^T\xi\mathcal{D}^in) \]\frac{\pd n}{\pd n^k}  ,
\end{align}
where $i\geq2$ and $1\leq m \leq i-1$. With the help of Eq.~(\ref{me2}) and after some algebra, one can show that $T_1=-T_2$ and $T_3=-T_4$. Therefore, we have
\begin{align}
-\frac{\pd\mathcal{H}_0}{\pd n^k} = L\mathcal{C}_k  ,
\end{align}
and
\begin{align}
\frac{\pd\mathcal{L}}{\pd n^k} = \mathcal{C}_i\[ L\delta^i_k + N\frac{\pd(\mathcal{D}^i{}_jn^j)}{\pd n^k} \].
\end{align}
Since the matrix $L\delta^i_k + N\frac{\pd(\mathcal{D}^i{}_jn^j)}{\pd n^k}$ is invertible, the $n^i$ equation of motion can be reduced to
\begin{align}
\mathcal{C}_i  = \mathcal{R}_i - \sqrt\gamma{n^l\xi_{lj} \over \sqrt x}\sum_{a=1}^{dC}\beta_a a(\sqrt x)^{a-1}\delta^j{}_{[i}\mathcal{D}^{p_1}{}_{p_1}\cdots \mathcal{D}^{p_{a-1}}{}_{p_{a-1}]}  =  0 .
\end{align}

\section{The Poisson brackets}

\subsection{Sector $(d-1)\Omega\neq-d\Omega^{,A}\Omega_{,A}$}
\label{sec:genpoissonb}

In this subsection, we calculate the Poisson brackets between $\mathcal{R}$ and $\mathcal{R}_i$ in the general sector and show, similar to General Relativity, they satisfy the following algebra,
\begin{align}
\{\mathcal{R}(x), \mathcal{R}(y)\}  &=-\left[\mathcal{R}^i(x)\pd_{x^i} \delta^d(x-y) - (x\leftrightarrow y)\right]  ,\nn
\{\mathcal{R}(x), \mathcal{R}_i(y)\}&=-\mathcal{R}(y)\pd_{x^i} \delta^d(x-y)  ,  \nn
\{\mathcal{R}_i(x), \mathcal{R}_j(y)\} &= -\[\mathcal{R}_j(x)\pd_{x^i} \delta^d(x-y) - \mathcal{R}_i(y)\pd_{y^j} \delta^d(x-y)\]  .
\end{align}
where $\mathcal{R}$ and $\mathcal{R}_i$ are defined in Eqs.~(\ref{Rexp}) and (\ref{Riexp}) respectively. We assume the reader is familiar to the relevant calculation in General Relativity which can be found in, for example,~\cite{Khoury:2011ay}.

To facilitate our calculation, we introduce time-independent smoothing functions $f(x)$, $f^i(x)$, $g^i(y)$ and $g(y)$ that decay quickly at infinities and define the smeared quantities
\begin{align}
F_R &\equiv \int\ud^dx~f(x)\mathcal{R}(x),& \quad F &\equiv \int\ud^dx~f^i(x)\mathcal{R}_i(x), \nn
G_R &\equiv \int\ud^dx~g(x)\mathcal{R}(x),& \quad G &\equiv \int\ud^dx~g^i(x)\mathcal{R}_i(x).
\end{align}
and will calculate the following Poisson brackets instead:
\begin{align}
\{ F_R , G_R \} &= \int \ud^dx \ud^dy f(x) g(y) \{\mathcal{R}(x), \mathcal{R}(y)\} ,\nn
\{ F_R , G \} &= \int \ud^dx \ud^dy f(x) g^i(y) \{\mathcal{R}(x), \mathcal{R}_i(y)\} ,\nn
\{ F , G \} &= \int \ud^dx \ud^dy f^i(x) g^j(y) \{\mathcal{R}_i(x), \mathcal{R}_j(y)\} \,.
\end{align}

\noindent$\bullet$ {$\{\mathcal{R}(x), \mathcal{R}(y)\}$}

Since $\{\mathcal{R}(x), \mathcal{R}(y)\}$ is antisymmetric in exchanging $x$ and $y$, we can drop terms proportional to $\delta^d(x-y)$ and expand it as
\begin{align}\label{r}
\{\mathcal{R}(x), \mathcal{R}(y)\}  = T_{\mathcal{R}\mathcal{R}1} + T_{\mathcal{R}\mathcal{R}2} +T_{\mathcal{R}\mathcal{R}3},
\end{align}
where
\begin{align}
T_{\mathcal{R}\mathcal{R}1}&=\bigg\{\sqrt\gamma{\Omega\over 2} {}^{(d)\!}R(x), -{1\over \sqrt\gamma}\[{2\over \Omega} \pi^{ij}\pi_{ij}
 - {1+\theta\over \omega}\pi^2 -  {1\over \omega} \pi\Omega_{,A}\pi^A\]\!(y)\bigg\} -(x\leftrightarrow y),
 \\
 T_{\mathcal{R}\mathcal{R}2}&=\bigg\{\sqrt\gamma D_kD^k\Omega(x), {1\over \sqrt\gamma}\bigg[ {2\over \Omega} \pi^{ij}\pi_{ij} + {1\over2} \pi^A \pi_A - {1+\theta \over \omega}\pi^2 - {1\over\omega} \pi\Omega_{,A}\pi^A  - {d \over 4\omega} (\Omega_{,A}\pi^A)^2 \bigg]\!(y)\bigg\}
 \nn
 &\qquad -(x\leftrightarrow y),
 \\
 T_{\mathcal{R}\mathcal{R}3}&=\bigg\{\sqrt\gamma {1\over 2}\pd_i \phi_B \pd^i \phi^B(x), {1\over \sqrt\gamma} \[ {1\over2} \pi^C \pi_C -  {1\over\omega} \pi \Omega_{,C}\pi^C - {d\over4\omega}{(\Omega_{,C}\pi^C)^2 \over \Omega} \]\!(y)\bigg\} -(x\leftrightarrow y)  .
\end{align}
We then simplify each of the terms above in turn.  Making use of the relation $\delta {}^{(d)\!}R = -\delta\gamma_{ij}{}^{(d)\!}R^{ij} + D^jD^i\delta\gamma_{ij} - D^kD_k\gamma^{ij}\delta\gamma_{ij}$ and dropping some boundary terms, the first term can be simplified to
\begin{align}\label{r11}
       &\int \ud^d x\ud^d y f(x)g(y)T_{\mathcal{R}\mathcal{R}1}
       \nn
=     & \int\ud^dx~\[ \({d-1\over2\omega} \Omega_{,A}\pi^A - {\theta\over\omega} \pi\) fD^k\Omega\pd_kg + {2\over\Omega}\pi_{mn}fD^n\Omega\pd^mg \]\!(x)\nn
      & + \int\ud^dy~\bigg[ g\Omega D_k\({d-1\over2\omega} \Omega_{,A}\pi^A - {\theta\over\omega}\pi\) \pd^kf + g\Omega D^m\({2\over\Omega}\pi_{mn}\)\pd^{y^n}f \bigg]\!(y) \nn
      &- (x\leftrightarrow y).
\end{align}
Inserting $f(y)=\int\ud^dx~f(x)\delta^d(x-y)$ and $g(x)=\int\ud^dy~g(y)\delta^d(x-y)$ into the expression above, we can arrive at
\begin{align}\label{r1}
       &\int \ud^d x\ud^d y f(x)g(y)T_{\mathcal{R}\mathcal{R}1}
       \nn
=      & \int\ud^dx\ud^dy~f(x)g(y) \bigg( \[ \({d-1\over2\omega} \Omega_{,A}\pi^A - {\theta \over \omega}\pi\) D^i\Omega + {2\over\Omega}\pi^{ij}D_j\Omega \]\!{(x)}\pd_{x^i}\delta^d(x-y) \nn
      & + \[ \Omega D^i\({d-1\over2\omega} \Omega_{,A}\pi^A - {\theta\over\omega}\pi\) + \Omega D_j\({2\over\Omega}\pi^{ij}\) \]\!{(y)}\pd_{y^i}\delta^d(x-y) \bigg) - (x\leftrightarrow y).
\end{align}
To simplify the second term in Eq.~(\ref{r}), we note that $\sqrt\gamma D_kD^k\Omega=\pd_k(\sqrt\gamma\gamma^{kl}\pd_l\Omega)$ and $\theta=\Omega^{,A}\Omega_{,A}/\Omega$. After some steps, we get
\begin{align}\label{r2}
       &\int \ud^d x\ud^d y f(x)g(y)T_{\mathcal{R}\mathcal{R}2}
       \nn
  =  & \int\ud^dx\ud^dy~f(x)g(y) \bigg[{\({d-2 \over 2\omega}\Omega_{,A}\pi^A - {1+2\theta \over \omega}\pi\)\pd^i\Omega }
      + {D^i\({\theta \over \omega}\Omega\pi-{d-1 \over 2\omega}\Omega\Omega_{,A}\pi^A\)}
      \nn
      &~~~ +{4\over\Omega}\pi^{ij}\pd_j\Omega \bigg]\!(y)\pd_{y^i}\delta^d(x-y) - (x\leftrightarrow y).
\end{align}
Then the last term  in Eq.~(\ref{r}) can be simplified to
\begin{align}\label{r3}
       &\int \ud^d x\ud^d y f(x)g(y)T_{\mathcal{R}\mathcal{R}2}
       \nn
  =  & \int\ud^dx\ud^dy  f(x)g(y) {\(-\pi_A\pd^i\phi^A + {1\over\omega}\pi\pd^i\Omega + {d \over 2\omega}\Omega_{,A}\pi^A \pd^i\Omega\)}\!{(y)} \pd_{y^i}\delta^d(x-y) \nn
   &~~~~~  - (x\leftrightarrow y).
\end{align}
Now, putting the three simplified terms together,  after some algebra (beware that $\pi$ is a scalar density rather than a scalar), we arrive at
\be
\{\mathcal{R}(x), \mathcal{R}(y)\}  =-\left[\mathcal{R}^i(x)\pd_{x^i} \delta^d(x-y) - (x\leftrightarrow y)\right].
\ee

\noindent$\bullet$ {$\{\mathcal{R}(x), \mathcal{R}_i(y)\}$}

First, we define
\begin{align}
F_{R}  &= F_{R1}+F_{R2}+F_{R3} \\
F_{R1} &= \int\ud^dx~\sqrt\gamma f(x){\Omega\over2}{}^{(d)\!}R \,,\\
F_{R2} &= -\int\ud^dx~\sqrt\gamma f(x)D_iD^i\Omega \,,\\
F_{R3} &= -\int\ud^dx~\bigg\{\sqrt\gamma f(x)\( {1\over2}\pd_i\phi_A\pd^i\phi^A+V \)+{f(x)\over\sqrt{\gamma}} \bigg[ {2\over\Omega} \pi^{ij}\pi_{ij} + {1\over2} \pi^A \pi_A \nn
& - {1+\theta\over \omega}\pi^2 -  {1\over\omega} \pi \Omega_{,A}\pi^A - {d\over4\omega} (\Omega_{,A}\pi^A)^2 \bigg]\bigg\} \,,
\end{align}
and calculate $\{F_{R1},G\}$, $\{F_{R2},G\}$ and $\{F_{R3},G\}$ in turn. The Poisson bracket $\{F_{R1},G\}$ contains two parts: the part involving variations on $\gamma_{mn}$ and its conjugate momenta, which is similar to the case of General Relativity, and the part involving the non-minimal coupling $\Omega$. Making use of the identities $D_iD_ng^i-D_nD_ig^i={}^{(d)\!}R_{in}g^i$ and $2\nabla_cR_a{}^c=\nabla_aR$, we can get
\begin{align}\label{fr1}
\{ F_{R1} , G \} =& \int\ud^dz~\bigg[ -{\sqrt\gamma\over2}\Omega f{}^{(d)\!}RD_ig^i - {\sqrt\gamma\over2}\Omega fg^iD_i{}^{(d)\!}R - {\sqrt\gamma\over2}f{}^{(d)\!}Rg^iD_i\Omega\bigg] \nn
               =& \int\ud^dz~fD_i\( -g^i\sqrt\gamma{\Omega\over2}{}^{(d)\!}R \) \,.
\end{align}
On the other hand, after dropping some boundary terms,  $\{ F_{R2} , G \}$ can be simplified to
\begin{align}
\{ F_{R2} , G \} = \int\ud^dz~\sqrt\gamma f\[ D_kD^k\Omega D_ig^i + g^iD_kD^kD_i\Omega - D^k\Omega(D_iD_kg^i - D_kD_ig^i) \] \,.
\end{align}
Making use of the identity $D_iD_ku^i-D_kD_iu^i={}^{(d)\!}R_{ik}u^i$, then we can get
\begin{align}\label{fr2}
\{ F_{R2} , G \}     = \int\ud^dz~fD_i\( g^i\sqrt\gamma D_kD^k\Omega \) \,.
\end{align}
Then, $\{ F_{R3} , G \}$, after several steps of straightforward simplification, can be written as
\begin{align}\label{fr3}
\{ F_{R3} , G \} &= \int\ud^dz~fD_i\bigg\{g^i\sqrt\gamma \( {1\over2}\pd_k\phi_A\pd^k\phi^A+V \)+{g^i\over\sqrt{\gamma}} \bigg[ {2\over\Omega} \pi^{kj}\pi_{kj} + {1\over2} \pi^A \pi_A \nn
& - {1+\theta \over \omega}\pi^2 -  {1\over\omega} \pi \Omega_{,A}\pi^A - {d \over 4\omega} (\Omega_{,A}\pi^A)^2 \bigg]\bigg\}.
\end{align}
Putting $\{F_{R1},G\}$, $\{F_{R2},G\}$ and $\{F_{R3},G\}$ together,  we have
\begin{align}
\{ F_R , G \} &= \int\ud^dx~fD_i\( -g^i\mathcal{R} \).
\end{align}
Notice that $D_i\( -g^i\mathcal{R} \)=\pd_i\( -g^i\mathcal{R} \)$ and $g^i(x)=\int\ud^dyg^i(y)\delta^d(x-y)$, we finally arrive at
\be
\{\mathcal{R}(x), \mathcal{R}_i(y)\} = -\mathcal{R}(y)\pd_{x^i}\delta^d(x-y) \,.
\ee

\noindent$\bullet$ {$\{\mathcal{R}_i(x), \mathcal{R}_j(y)\}$}

We first define
\begin{align}
\{ F , G \}     &=\{ F , G \}_1 + \{ F , G \}_2,
\\
\{ F , G \}_1 &= \int\ud^dz~\({\delta F\over\delta\gamma_{mn}}{\delta G\over\delta\pi^{mn}} - {\delta F\over\delta\pi^{mn}}{\delta G\over\delta\gamma_{mn}}\),
\\
\{ F , G \}_2 &= \int\ud^dz~\({\delta F\over\delta\phi_A}{\delta G\over\delta\pi^A} - {\delta F\over\delta\pi^A}{\delta G\over\delta\phi_A}\).
\end{align}
Calculation of $\{ F , G \}_1$ is similar to the case of General Relativity and the result is
\be
\{ F , G \}_1 = -\int\ud^dx\ud^dy~f^i(x)g^j(y)\bigg[ {2\gamma_{jk}D_l\pi^{kl}}(x)\pd_{x^i}\delta^d(x-y) - {2\gamma_{ik}D_l\pi^{kl}}{(y)}\pd_{y^j}\delta^d(x-y) \bigg] \,.
\ee
Also, the term $\{ F , G \}_2$ is straightforward to calculate and we have
\begin{align}
\{ F , G \}_2 = \int\ud^dz~\[ f^i\pi^A\pd_j\phi_A\pd_ig^j - g^j\pi^A\pd_i\phi_A\pd_jf^i \].
\end{align}
Adding them together and inserting $g^j(x)=\int\ud^d yg^j(y)\delta^d(x-y)$ and $f^i(y)=\int\ud^d x f^i(x)\delta^d(x-y)$, we have
\be
\{\mathcal{R}_i(x), \mathcal{R}_j(y)\} = -\[\mathcal{R}_j(x)\pd_{x^i} \delta^d(x-y) - \mathcal{R}_i(y)\pd_{y^j} \delta^d(x-y)\].
\ee

\subsection{Sector $(d-1)\Omega=-d\Omega^{,A}\Omega_{,A}$}
\label{sec:appsc}

In this subsection, we calculate the Poisson brackets between $\mathcal{R}$, $\mathcal{R}_i$ and $\mathcal{C}_\lambda$ for the sector $(d-1)\Omega=-d\Omega^{,A}\Omega_{,A}$. We show that they satisfy the following relations
\begin{align}
\{\mathcal{R}(x), \mathcal{R}_i(y)\}&=-\mathcal{R}(y)\pd_{x^i} \delta^d(x-y) ,
\\
\{\mathcal{R}_i(x), \mathcal{R}_j(y)\} &= -\[\mathcal{R}_j(x)\pd_{x^i} \delta^d(x-y) - \mathcal{R}_i(y)\pd_{y^j} \delta^d(x-y)\] ,
\\
\{\mathcal{R}_i(x), \mathcal{C}_\lambda(y)\}&=-\mathcal{C}_\lambda(x)\pd_{x^i}\delta^d(x-y) ,
\\
\{\mathcal{C}_\lambda(x), \mathcal{C}_\lambda(y)\}&=0  ,
\\
\label{App2RR}
\{\mathcal{R}(x),\mathcal{R}(y)\} &=-\[(\mathcal{R}^i-\frac2d D^i\mathcal{C}_{\lambda})(x)\pd_{x^i} \delta^d(x-y) - (x\leftrightarrow y)\] ,\\
\label{App2RC}
\{\mathcal{R}(x), \mathcal{C}_\lambda(y)\} &=\bigg[ \sqrt\gamma\bigg( -{\Omega\over4}{}^{(d)\!}R - {d-1\over2}D^kD_k\Omega + \frac d2\Omega_{,A}D_kD^k\phi^A - {d-2\over4} \pd_k\phi_A\pd^k\phi^A - \frac d2 V \nn
 &- \frac d2 V^{,A}\Omega_{,A} \bigg) + {1\over\sqrt\gamma}\( {1\over\Omega}\pi^{kj}\pi_{kj} + \frac d4\pi^A\pi_A + \frac d4 {(\Omega_{,A}\pi^A)^2\over\Omega} - {1\over\Omega d}\pi^2 \)  \bigg](x)\delta^d(x-y) ,
\end{align}
where $\mathcal{C}_\lambda=\pi+d \Omega_{,A}\pi^A/2$, and $\mathcal{R}$ and $\mathcal{R}_i$ are respectively defined by taking the $\theta\to \infty$ limit of  the expressions $\mathcal{R}$ (Eq.~(\ref{Rexp})) and $\mathcal{R}_i$ (Eq.~(\ref{Riexp})) in the general sector. 
Note that, for the case of one scalar, if $2({d+1})V = (d-1)(\phi+c)V'$ is satisfied ($V(\phi)=c_1(\phi+c)^{2(d+1)/(d-1)}$, $c_1$ being a integration constant),  we get a closed Poisson algebra for $\mathcal{R}$, $\mathcal{R}_i$ and $\mathcal{C}_\lambda$.\;\footnote{If $2({d+1})V = (d-1)(\phi+c)V'$ is satisfied, the enhanced symmetry for the case of a zero graviton mass is diffeomorphism invariance plus a conformation invariance (see \cite{Sotiriou:2006hs}).}  The calculations of $\{\mathcal{R}(x), \mathcal{R}_i(y)\}$ and $\{\mathcal{R}_i(x), \mathcal{R}_j(y)\}$ are essentially the same as in the general sector case and the calculations of $\{\mathcal{R}_i(x), \mathcal{C}_\lambda(y)\}$ and $\{\mathcal{C}_\lambda(x), \mathcal{C}_\lambda(y)\}$ are also straightforward. We assume the reader is familiar with Appendix \ref{sec:genpoissonb}, and will focus on the later two Poisson brackets and keep the discussion brief.

As in the general sector case, we divide $\{\mathcal{R}(x),\mathcal{R}(y)\}$ into three parts:
\begin{align}\label{rnew}
\{\mathcal{R}(x), \mathcal{R}(y)\} &=T_{\mathcal{R}\mathcal{R}1}+T_{\mathcal{R}\mathcal{R}2}+T_{\mathcal{R}\mathcal{R}3},
\\
T_{\mathcal{R}\mathcal{R}1}&=\bigg\{\sqrt\gamma{\Omega\over 2} {}^{(d)\!}R, -{1\over \sqrt\gamma}\[{2\over \Omega} \pi^{ij}\pi_{ij} - \frac{2}{\Omega d} \pi^2 \]\bigg\}
-(x\leftrightarrow y),
\\
 T_{\mathcal{R}\mathcal{R}2}&=\bigg\{-\sqrt\gamma D_kD^k\Omega, -{1\over \sqrt\gamma}\bigg[ {2\over \Omega} \pi^{ij}\pi_{ij} + \frac12 \pi^A\pi_A - \frac{2}{\Omega d} \pi^2 \bigg]\bigg\}
 -(x\leftrightarrow y),
 \\
 T_{\mathcal{R}\mathcal{R}3}&=\bigg\{-\sqrt\gamma {1\over 2}\pd_i \phi_B \pd^i \phi^B, -{1\over \sqrt\gamma} \frac12 \pi^A\pi_A \bigg\} -(x\leftrightarrow y).
\end{align}
Going through similar steps, we get
\begin{align}
T_{\mathcal{R}\mathcal{R}1}
&=\int\ud^dx\( \frac{4}{\Omega}\pi^{ij}D_j\Omega - \frac{4}{\Omega d}\pi D^i\Omega - 2D_j\pi^{ij} + \frac2dD^i\pi \)(f\pd_ig-g\pd_if),
\\
T_{\mathcal{R}\mathcal{R}2}&=\int\ud^dx\[ -\frac{4}{\Omega}\pi^{ij}D_j\Omega + \frac{4}{\Omega d}\pi D^i\Omega + D^i(\Omega_A\pi^A) \](f\pd_ig-g\pd_if),
\\
T_{\mathcal{R}\mathcal{R}3}&=\int\ud^dx\pi_A\pd^i\phi^A(f\pd_ig-g\pd_if) \,.
\end{align}
Adding them together, integrating by parts and recognizing $\pi+{d\over2}\Omega_{,A}\pi^A=\mathcal{C}_\lambda$, we can get Eq.~(\ref{App2RR}). It is different from the general sector case because for $\{\mathcal{R}(x),\mathcal{R}(y)\}$ evaluating the Poisson bracket and taking the limit $\theta\rightarrow\infty$ do not commute. More specifically, the term $\{ \sqrt\gamma D_kD^k\Omega(x), -\frac{1}{\sqrt\gamma}\[\frac{\pi}{\omega}\pi\Omega_{,A}\pi^A+\frac{d}{4\omega}(\Omega_{,A}\pi^A)^2\](y) \}-(x\leftrightarrow y)$ gives rise to $-\[ \frac2d D^i\mathcal{C}_\lambda(x)\pd_{x^i}\delta^d(x-y)-(x\leftrightarrow y) \]$ if the limit $\theta\rightarrow\infty$ is taken after evaluating the Poisson bracket, but does not contribute if the limit is taken first.

For the Poisson bracket $\{\mathcal{R}(x), \mathcal{C}_\lambda(y)\}$, we can expand it out to four terms, which can be simplified, using the smoothing function technique, to
\begin{align}
\int\ud^d x\ud^d yf(x)g(y)\{\mathcal{R}(x),\mathcal{C}_\lambda(y)\} &=T_{\mathcal{R}\lambda1}+T_{\mathcal{R}\lambda2}+T_{\mathcal{R}\lambda3}+T_{\mathcal{R}\lambda4},
\end{align}
where
\begin{align}
T_{\mathcal{R}\lambda1}
&= \int \ud^dz \( -{f\lambda\over4}\sqrt\gamma{\Omega}{}^{(d)\!}R  - {d-1\over2}\lambda\sqrt\gamma D^kD_k(\Omega f)  \),
\\
T_{\mathcal{R}\lambda2}&=\int \ud^dz \( -f\sqrt\gamma{d-2\over2}D_k(\lambda D^k\Omega) + {d-1\over2}\lambda\sqrt\gamma\Omega D^kD_kf \),
\\
T_{\mathcal{R}\lambda3}&=\int \ud^dz \left[-f\lambda\sqrt\gamma{d-2\over4}\pd_k\phi_A\pd^k\phi^A + {\lambda\sqrt\gamma d\over2}\Omega_{,A}D_k(fD^k\phi^A) \right.
\nn
&~~~~~~\left. - f\lambda\sqrt\gamma\frac d2V - f\lambda\sqrt\gamma\frac d2V^{,A}\Omega_{,A} \right],
\\
T_{\mathcal{R}\lambda4}&= \int \ud^dz {f\lambda\over2\sqrt\gamma}\[ {2\over\Omega}\pi^{kj}\pi_{kj} + {d\over2}\pi^A\pi_A - {2\over\Omega d}\pi^2 + {d\over2} \frac{(\Omega_{,A}\pi^A)^2}{\Omega} \].
\end{align}
Combining them together, after some algebra, we get Eq.~(\ref{App2RC}).

\end{document}